# Soft capacitor fibers using conductive polymers for electronic textiles

Jian Feng Gu, Stephan Gorgutsa, Maksim Skorobogatiy

*Abstract*— A novel, highly flexible, conductive polymer-based fiber with high electric capacitance is reported. In its crossection the fiber features a periodic sequence of hundreds of conductive and isolating plastic layers positioned around metallic electrodes. The fiber is fabricated using fiber drawing method, where a multi-material macroscopic preform is drawn into a sub-millimeter capacitor fiber in a single fabrication step. Several kilometres of fibers can be obtained from a single preform with fiber diameters ranging between 500μm -1000μm. A typical measured capacitance of our fibers is 60-100 nF/m and it is independent of the fiber diameter. For comparison, a coaxial cable of the comparable dimensions would have only ~0.06nF/m capacitance. Analysis of the fiber frequency response shows that in its simplest interrogation mode the capacitor fiber has a transverse resistance of 5 kΩm/L, which is inversely proportional to the fiber length L and is independent of the fiber diameter. Softness of the fiber materials, absence of liquid electrolyte in the fiber structure, ease of scalability to large production volumes, and high capacitance of our fibers make them interesting for various smart textile applications ranging from distributed sensing to energy storage.

*Index Terms*—Polymers, Flexible structures, Capacitors, Fibers, Conducting films, Dielectric films, Energy storage, E-textiles, Smart textiles

## I. INTRODUCTION

Fuelled by the rapid development of micro and nanotechnologies, and driven by the need to increase the value of conventional textile products, fundamental and applied research into smart textiles (or high-tech textiles) has recently flourished. Generally speaking, textiles are defined as "smart" if they can sense and respond to the environmental stimuli that can be of mechanical, thermal, chemical, electrical, magnetic, etc. nature. Some of the first uses of smart textiles were in military and medical applications. For example, with the support of US Naval Department in 1996 Georgia Tech has developed a garment called Wearable Motherboard (with the commercial name of Smart shirt) [1], [2]. The Wearable Motherboard is a fabric featuring woven electric wires and/or optical fibres that serve as a flexible information bus.To integrate electronics directly into textiles leads to the so called technique of "wearable computing" or "E-textiles"[3]-[5]. Another application of smart textiles is in harnessing (and recently in storage) of the energy of human motion or the energy of various ambient fields, such as electromagnetic fields. Electric energy generation from human motion, for example, has been recently demonstrated using piezoelectric fibers made of ceramic materials like PZT (lead zirconate-titanate) as well as polymers such as PVDF (polyvinylidene fluoride) [6],[7]. Smart textiles can also be found their usage in heat-storage and thermo-regulated clothing [8],[9] and various wearable sensors including those for biomedical monitoring [10]. For example, conventional fabrics coated with a thin conducting polymer layer possess remarkable properties of strain and temperature sensing [11]. A multi-layer structure consisting of two conductive fabrics separated by a meshed non-conductive one can be used as a pressure sensor [10]. Sensing garments for monitoring physiological and biomechanical signal of human body have already been invented for healthcare [12] and sports training [13]. Other applications of smart textiles have been demonstrated from responsive seats in automobiles [14] where textiles can indicate the level of comfort of an individual passenger, to apparel with tuneable or adjustable color and appearance in fashion and design [15]. With the constant improvement of the technology, there is no doubt that smart textiles will soon become an integral part of our daily life [16]-[18].

Most of the "smart" functionalities in the currently existing smart apparel are enabled by various point devices attached to a textile matrix. Such point devices (electronic chip-sets, for example) are typically not compatible with the weaving process and, thus, have to be introduced onto a textile surface in a post-processing step. This, in turn, makes fabrication of smart apparel labour intensive and therefore expensive. Additionally, most of the point devices such as sensors and batteries are not flexible, and therefore wearability of such garments is usually problematic. Limitations introduced by the rigid point devices motivated recent effort into the development of flexible electronic components. For example, a flexible energy

All the authors are with Ecole Polytechnique de Montréal, C.P. 6079, succ. Centre-ville, Montréal (Québec), Canada  H3C 3A7 (tel: 514-340-4711-3327 ; fax: 514-340-3218; e-mail: maksim.skorobogatiy@ polymtl.ca).

storage devices based on nanocomposite paper was reported [19] . The device, engineered to function as both a lithium-ion battery and a supercapacitor, can provide the long and steady power output.

Ideally, if all the electronic functionalities could be realised in a fiber itself, such fibers would provide a perfect building material for smart apparel as they could be naturally integrated into textiles during weaving process. Because of the technical complexity of integration of advanced electronic functionalities into a textile fiber, currently there are only few examples of such fibers. Thus, the researcher group of Wang [20] has developed a microfiber nanogenerator composed of a pair of entangled fibers which can generate electrical current using the piezoelectric effect. It has been demonstrated that commodity cotton threads can be transformed into smart electronic yarns and wearable fabrics for human biomonitoring using a polyelectrolyte-based coating with carbon nanotubes (CNTs) [21]. A stretchable, porous, and conductive textiles have recently been invented by a simple "dipping and drying" process using a single-walled carbon nanotube (SWNT) ink [22] . The loading of pseudocapacitor materials into these conductive textiles can lead to a 24-fold increase of the areal capacitance of the device. Finally, several groups [23]-[25] have recently demonstrated organic all-fiber transistors which can potentially allow creation of electronic logic circuits by weaving.

Recently there have been several reports on the capacitor fibers compatible with a textile weaving process. One possible application of a capacitor fiber is in distributed sensing of electric influence, proximity, etc. By adding an external inductance such fibers make a resonant LC circuit, thus allowing the use of many highly sensitive resonant detection techniques which are able to detect small changes in the capacitor structure. Probably more interestingly, capacitor fibers together with energy harvesting fibers, promise an all-textile solution for the problem of wearable energy generation and storage, with the principal advantage of a capacitor over an electrochemical battery being a capacitor almost unlimited cycle life. Among several proposals for a capacitor fiber we note a multicore fiber capacitor in the form of a bundle of ~50μm-sized coaxial cables connected in parallel on a micro-level (see Cheng and Hong [26]). However, such capacitors were never actually fabricated. Load bearing composite textiles comprising a large number of simple coaxial cables have been recently reported by the Air Force Research Laboratories [27] for distributed storage of electrical energy directly in the fuselage of an airplane for pulsed weapons applications. The invention envisions reduction in the aircraft payload by combining mechanical and electric functionalities in the same fiber. Finally, the group of Baughman [28] proposed carbon nanotube-based textile threads that can be adapted to build supercapacitor textiles after soaking such threads in electrolite.

In this paper we present a novel type of electronic fiber recently developed in our laboratory - high capacitance, soft fiber from conductive polymers. One key advantage of our fibers is that they do not require the use of electrolytes for their operation, which is especially desirable for wearable applications. Another key advantage is that the fibers can be made fully polymeric (no metallic electrodes) and very soft for applications in wearable sensing. Because of the fiber relatively high capacitance (60-100nF/m) it can be also used for energy storage applications. In terms of capacitance out fibers take an intermediate position between the coaxial cables and supercapacitors. Thus, capacitance of a coaxial cable with comparable parameters is typically 1000 times smaller than that of our fibers.

## II. FIBER CAPACITOR MATERIALS

The fiber capacitors presented in our paper are fabricated by the fiber drawing technique which consists of three steps. The first step involves rolling or stacking conductive and dielectric films into a multilayer preform structure. During the second step the preform is consolidated by heating it to temperatures somewhat above the polymer glass transition temperature ($T_g$). Finally, the third step involved drawing of the consolidated preform into fibers using fiber drawing tower. Drawing is typically performed at temperatures higher than the polymer $T_g$. During successful drawing the resultant fibers generally preserve the structured profile of a preform, thus, fibers with very complex microstructure can be fabricated via homologous reduction (during drawing) of a macrostructure of the preform. Drawing technique used in this paper is directly analogous to the one used in manufacturing of microstructured polymer optical fibers [29].

Flexible multilayer capacitors discussed in this paper generally consist of two conducting polymer layers serving as two electrodes of a capacitor, and two isolating polymer separator layers. To result in a successful drawing the perform materials should be compatible with each other in terms of their various rheological and thermo-mechanical properties. In our first tests we have attempted drawing of a thin continuous layer of low-melting temperature metal sandwiched between the two isolating polymer layers. As a metal we have used Bi58/Sn42 alloy with a melting point of 138°C. Various polymers were tested as isolating layers. However, we found that in all cases it was difficult to preserve the laminated structure with a thin metal foil during drawing process. Particularly, when melted, metal foil would break into wires during drawing, thus destroying the continuous electrode structure. We have rationalised this observation by noting that the viscosity of melted metal is very low compared to that of a surrounding plastic, thus it is easy for a thin sheet of melted metal to develop a flow instability and form several larger wires to minimise surface energy associated with a polymer/metal interface. Another potential problem during drawing of metal sheets is that when the polymer surrounding the molten metal becomes too soft it can no longer hold the melt; as a consequence a large drop of metal would form at the perform end even before drawing starts, thus draining the rest of a perform from metal. From these initial experiments we have concluded that drawing of a thin metallic sheet sandwiched between two plastic sheets is in general problematic due to strong mismatch of the material viscosities during drawing process. In principle, one can remedy this situation by using amorphous metals with viscosities that exhibit continuous rather than a rapid change in the temperature range of drawing. However, such metals are still quite exotic and their thermal processing is not trivial.

After realising the challenge of drawing metallic electrods in the form of thin sheets, a natural option to remedy this problem was to substitute metals with thermoelastic conductive polymers as electrodes. Unfortunately, thermoplastic intrinsic conductive polymers are not available commercially. The only thermoplastic conductive polymers which are currently available commercially are either carbon black filled or most recently carbon nanotube filled films. In our research we have mostly used polyethylene-based carbon black filled films (BPQ series) provided by Bystat International Inc. The film has a surface resistivity of 17 KΩ/sq. To find the isolating materials that can be co-drawn with this conductive film, we have tried various polymer films such as polyvinylidene fluoride (PVDF), polycarbonate (PC), polyethylene terephthalate (PET), polymethyl methacrylate (PMMA), and others. Among all the attempted polymers we have found that the two best materials for the isolating layer were low density polyethylene (LDPE) film or a polycarbonate film. The fibers were either drawn bare or they were encapsulated into electrically isolating PMMA jacket.

Finally, to connectorize drawn fibers several approaches were attempted. In one implementation the fiber only contained conductive polymer layers and to interrogate such fibers one had to introduce external probes into the fiber structure. In the second implementation either a small diameter copper wire was integrated into fiber during drawing, or a tin alloy electrode was drawn directly during fiber fabrication, thus providing a convenient way to connect one of the two probes. In the third implementation, two small copper wires were integrated during drawing, thus providing a convenient way to attach both electrical probes.

## III.  CAPACITOR FIBER DESIGNS

Three distinct fiber capacitor geometries were successfully explored. The first fiber type features cylindrical geometry with two plastic electrodes in the form of a spiralling multilayer (see Fig. 1(a)). Central part of a fiber was either left empty with the inner plastic electrode lining up the hollow core, or a metallic electrode was introduced into the hollow core during drawing, or the core was collapsed completely thus forming a plastic central electrode. In all these fibers, the second electrode was wrapped around the fiber. The second fiber type (see Fig. 1(b)) is also of cylindrical multilayer geometry, however it features two hollow cores lined with two plastic electrodes. The fiber is wrapped into an isolating material so there is no direct contact with environment. During drawing two metallic electrodes were introduced into the fiber cores. Finally, the third fiber type features a square electrically isolating tube comprising a zigzagging stack of the plastic electrodes (see Fig. 1(c)) separated by a zigzaging dielectic layer. The metallic electrodes were integrated on the left and right sides of a tube for the ease of connectorization.
Fibers of the first and second types (Fig. 1(a)) were fabricated by co-rolling of the two conductive polymer films which were physically and electrically separated with the two isolating LDPE films. In the resultant fiber the inner conductive film forms one electrode inside the hollow fiber core, while another electrode is created by the other conductive film at the fiber surface. The preform was thermal consolidated at 105°C for one hour, and then drawn in a drawing tower at temperatures in the range of 170-185°C. Similar fabrication strategy was used in the fabrication of the second fiber type (Fig. 1(b)), with the only exception of positioning two isolated fiber cores in the fiber center, while encapsulating the fiber into an isolating HDPE plastic wrap. Finally, fibers of the third type (Fig. 1(c)) were created by encapsulating a zigzagging stack of the electrodes and isolating layers inside a rectangular PMMA tube. After consolidation at 135°C for one hour, the preform was drawn at temperatures around 200°C.

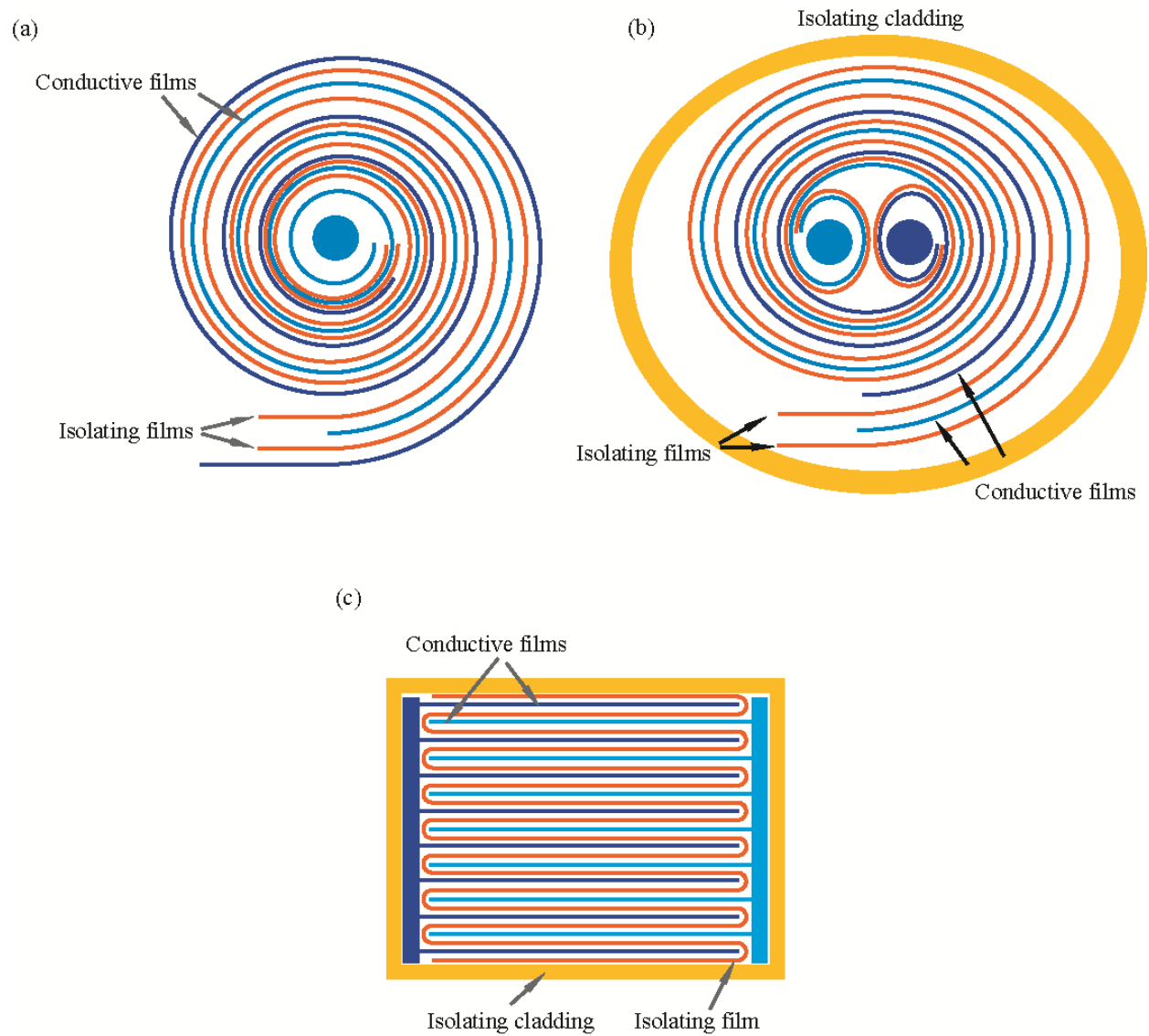

Fig. 1. (a) Schematic of a cylindrical capacitor fiber preform featuring a spiralling multilayer comprising two conductive and two isolating films. The deep blue and light blue curves represent two conductive films, while red curves represent isolating LDPE films. (b) Schematic of a cylindrical capacitor fiber with two electrodes in the center. (c) Schematic of a rectangular preform prepared by encapsulating a zigzaging stack of two conductive and an isolating layer inside a rectangular PMMA tube. The deep blue and light blue curves represent conductive films, while the red curves represent isolating polycarbonate (PC) films.

## IV. CAPACITOR FIBER CONNECTORIZATION, AND POTENTIAL APPLICATIONS

An important issue when designing any smart fiber concerns connectorization of such fibers either to each other or with the external electrical probes. In a view of various potential applications of a capacitor fiber we have explored several connectorization geometries. In Fig. 2(a)-(e) we show four complete designs for a capacitor fiber. In each figure we present both the structure of a preform before drawing, as well as the structure of a resultant fiber.

Design one is presented in Fig. 2(a) and 2(b) where we show a circular hollow core fiber with the first electrode formed by the conductive layer lining the hollow fiber core (see also Fig. 1(a)), and the second electrode formed by the other conductive layer wrapping the fiber from outside. The outside electrode is exposed for the ease of access. Hollow core can be either collapsed (Fig. 2(b)) or left open during drawing (Fig. 2(a)). In general, to access the electrode inside the fiber core one has to use a needle-like electrical probe; in fact, we have used 50μm-100μm diameter hypodermic needles to perform electrical characterization of this fiber. One of the advantages of the hollow core fibers is that they are very soft due to the lack of metallic components in their structure, and, therefore, are most suitable for the integration into wearable textiles. Moreover, hollow fiber core can be filled with functional liquids which will be in direct contact with one of the electrodes. This can be useful for various sensing applications, where physical or chemical properties of a liquid could be interrogated electrically.

Design two is presented in Fig. 2© where we show a circular hollow core fiber with one of the electrodes formed by a small 100μm diameter copper wire which is integrated into the fiber core directly during drawing. With a tension-adjustable reel installed on the top of a preform, copper wire can be passed through the perform core, pulled down and embedded into the fiber center during drawing by collapsing plastic cladding around it. The second electrode is formed by the other conductive layer wrapping the fiber from outside, similarly to the first design. Main advantage of this design is the ease of connectorization to the inner electrode as plastic capacitor multilayer can be easily stripped from the copper wire. This fiber has lower effective resistivity compared to the hollow core fiber as one of the electrodes is made of a highly conductive metal. Despite of the copper electrode in its structure the fiber is still highly flexible. As outside electrode is exposed, this fiber can be used for the detection of electromagnetic influence or as a proximity sensor.

Design three is presented in Fig. 2(d), where we show a circular fiber containing two hollow cores positioned in the middle of the fiber. Each core is lined with a distinct conductive layer which are forming electrode one and two. The cores with electrodes are electrically isolated from each other. Moreover, the whole preform is then wrapped into several layers of pure HDPE plastic to isolate the capacitor layers from the environment. The preform is then drawn with two copper wires threaded through its holes. Resultant fiber features two copper electrodes and a fully encapsulated capacitor multilayer. Such fibers can be interesting for energy storage applications due to ease of connectorization and electrical isolation from the environment.

Finally, design four is presented in Fig. 2(e) where we show a thin PMMA tube of square crossection comprising a zigzagging multilayer of the two conductive layers separated by a single electrically isolating PC layer. The first plastic electrode is located to the left and the second plastic electrode is located to the right of the isolating PC layer. At the left and right inner sides of the square tube we place foils of Bi58/Sn42 alloy in contact with the plastic conductive layers. During fiber drawing wire-like metallic electrodes are created from the foils. Finally, the structure of the resultant fiber is similar to the one of an encapsulated fiber with two copper electrodes.

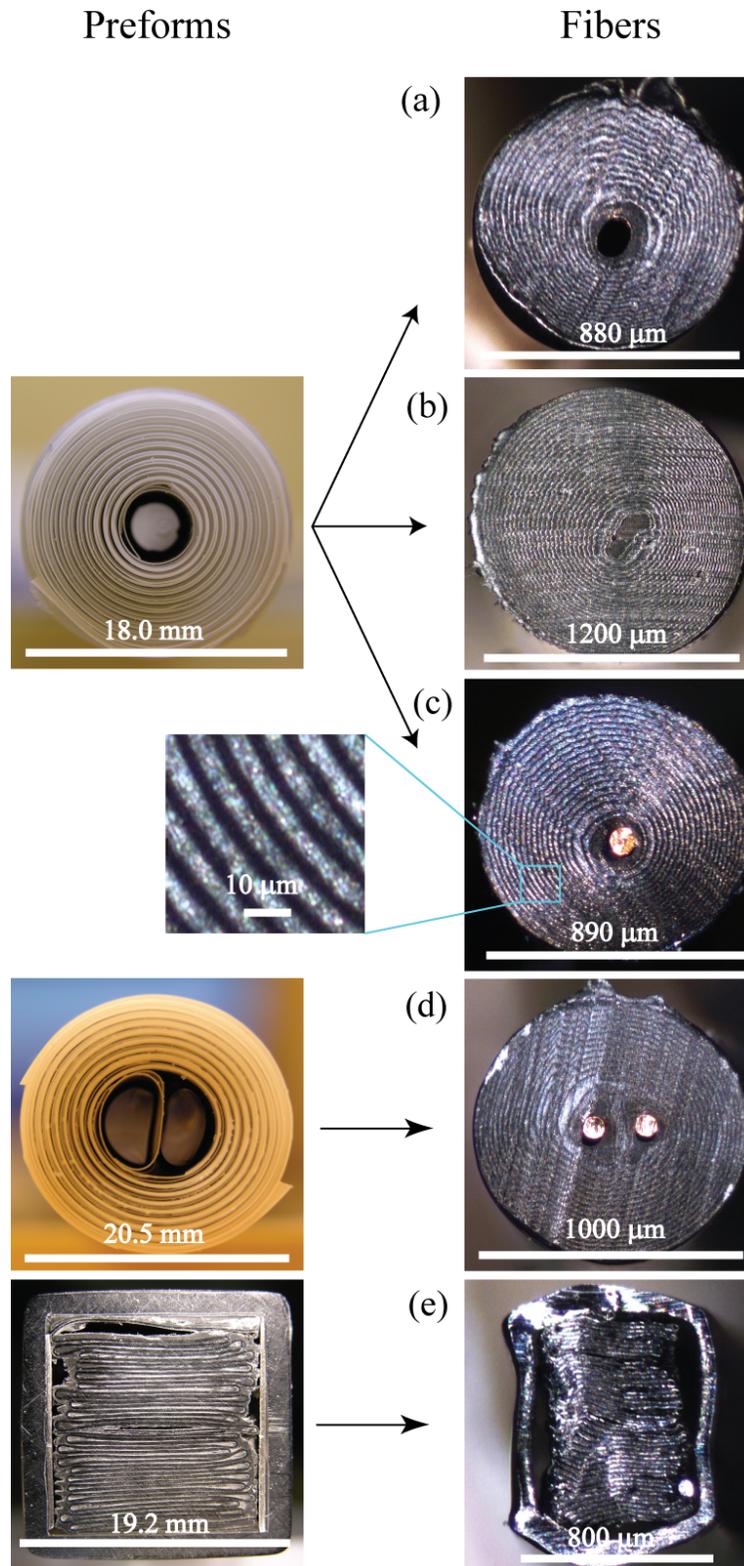

Fig. 2. Design I: hollow core fiber with the first electrode lining the inside of a holow core, and the second plastic electrode wrapping the fiber from outside. During drawing fiber hollow core can be left open (a) or collapsed (b) depending on the application requirement. Design II: Hollow core fiber can be drawn with a metallic electrode in the center. Such an electrode can be a copper wire (c) in contact with the plastic electrode lining the hollow core. Design III: fiber containing two hollow cores. The cores are lined with two plastic electrodes electrically separated from each other. Fiber is drawn with two copper wires threaded through the hollow cores in the preform (d). Design IV: square fiber capacitor. Fiber features a zigzaging stack of two plastic electrodes separated with an electrically isolating PC layer (e). Furthermore, two metallic electrodes are placed in contact with plastic electrodes, and the whole multilayer is encapsulated inside a square PMMA tube.

## V. MEASUREMENT OF ELECTRICAL CHARACTERIZATION OF CAPACITOR FIBERS

To characterize electrical properties of our capacitor fibers we used a measurement circuit presented in Fig. 3, where the fiber capacitor is connected to a function generator (GFG-8216A, Good Will Instrument Co., Ltd) through the reference resistor $R_{Ref}$= 480 kΩ. The function generator provides a sinusoidal signal of tuneable frequency ω=[0.3Hz-3MHz]. An oscilloscope (GDS-1022, Good Will Instrument Co., Ltd) measures the input voltage $V_{Ch1}$ (ω) on channel 1 and the output voltage over the reference resistor $V_{Ch2}$ (ω) on channel 2. A 10X probe (GTP-060A-4, Good Will Instrument Co., Ltd) was used to acquire the experimental data. The voltage produced by the function generator is fixed and in the whole frequency range of interest equals to $V_{FG}$ =2V. In our experiments we measure both the amplitudes and the phase difference between channels 1 and 2. Due to high resistivity of our fibers and also to fit the experimental data at higher frequencies (ω>1kHz), we have to take into account the effective impedances of a function generator and an oscilloscope.

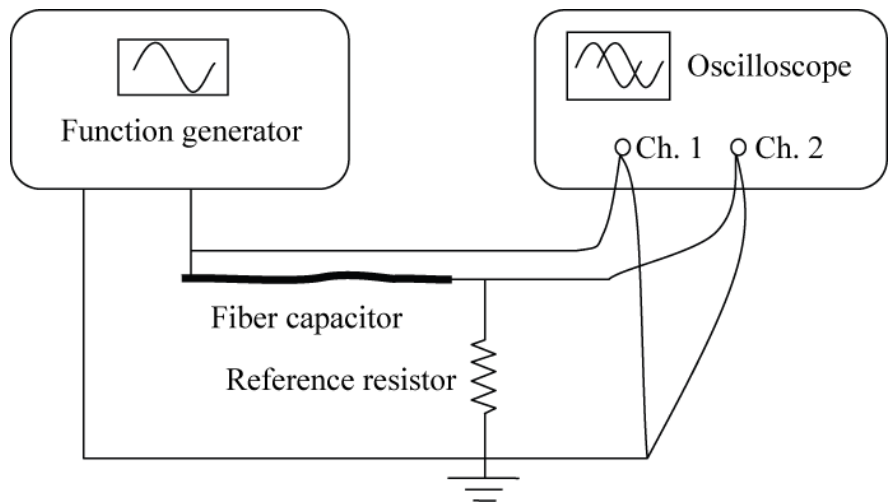

Fig. 3. Schematic of a measurement setup

To characterize capacitor fiber properties, we assume that it can be represented by an ideal capacitor with capacitance $C_F$ connected in series to an equivalent resistor with resistance $R_F$. This assumption is strictly valid when the fiber is uniform and when two electrodes implanted into or wrapped around the capacitance fiber are of low resistance (meaning that there is no voltage differential along the fiber length).

Before analysing our capacitor fibers we perform a calibration measurement to find the effective circuit parameters of an oscilloscope and a function generator. Calibration circuit is identical to that shown in Fig. 3 with the only exception being that instead of a fiber we use a known resistor $R_F$ =479 kΩ and no capacitance $C_F$. Now we use the developed effective circuit model to characterize fiber capacitors. As an example, in Fig. 4 we present electrical response of a 650μm diameter, 137 mm long capacitor fiber. On Fig. 4(a) we present $V_{Ch2}/ V_{Ch1}$ as a function of frequency and on Fig. 4(b) we present the phase difference between two channels also as a function of frequency. The fitting of data sets in the region of frequencies $\omega \sim 1Hz - 10kHz$ gives $C_F$ =9.8 nF and $R_F$ =26 kΩ.

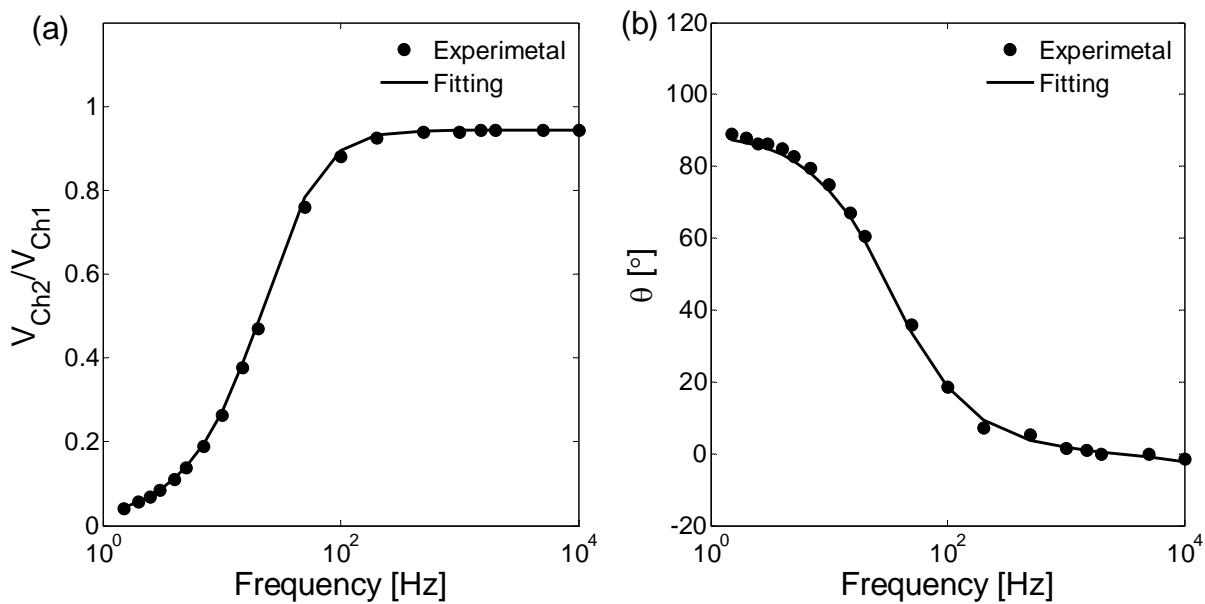

Fig. 4. Example of fitting of the capacitor fiber parameters. Measure fiber is 650μm in diameter and 137 mm in length. (a) and (b) represent the voltage drop and phase shift respectively versus frequency. Experimental data is shown as dots, while solid curves present a response of the effective circuit model with the fitted parameters.

## VI. ELECTRICAL PROPERTIES OF CAPACITOR FIBERS

In this section we present the properties of fiber capacitance and resistance as a function of various fiber geometrical parameters. Most of the measurements presented in this section were performed on fibers featuring a single copper electrode in their cores, while the second electrode is formed by the plastic conductive layer on the fiber surface (see Fig. 2(c)). The fiber was co-drawn with a 100μm-thick copper wire in its center. In the preform, both conductive layers are 75μm thick, while the two insulating layers are made of 86μm thick LDPE films. Examples of drawn fibers of ~1mm diameter are presented in Fig. 5(a). To characterise capacitance fibers we used embedded copper wire as the first electrical probe, while the second electrical probe was made by wrapping aluminum foil around a part or the whole of the fiber as shown in Fig. 5(b).

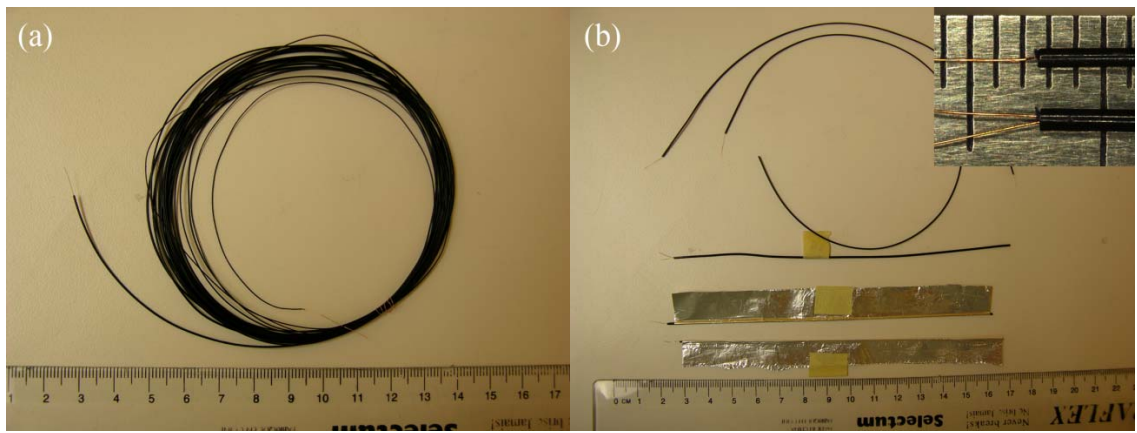

Fig. 5. (a) Capacitor fiber fabricated from the preform shown in Fig. 2(c). The fiber features a central 100μm-thick copper wire, as well as an exposed conductive plastic electrode on the fiber surface. (b) To perform electrical characterization of the fibers, embedded copper wire is used as the first electrical probe, while the second electrical probe is an aluminum foil wrapped around the fiber conductive surface. The inset is an enlarged view of the fibers with single and double copper wire electrodes.

### A. Effect of the Shape of an Electrical Probe Connected to the Fiber Outer Electrode

To connect the capacitor fibers to a measurement circuit one has to use two electrical probes. One of the probes is naturally a copper wire going through the fiber center. The other probe has to be attached to the surface of the fiber outer plastic electrode.

This electrode is quite resistive as it is made of a conductive plastic, therefore, measured results will be strongly influenced by geometry of the second probe. If a point probe is used to connect to the fiber outer electrode the measured resistance will be the highest, while if a distributed probe is used then the resistivity will be the smallest. This is easy to rationalise by noting that generally there will be two types of currents flowing through the fiber. The first current is transverse to the fiber direction, while the second one is along the length of the fiber. For the transverse current resistivity of a fiber will be:

$$R_F^t \propto \rho_v \frac{ND_F}{Ld_c} \quad (1),$$

where $N$ is the number of conductive bi-layers in a fiber, $\rho_v$ is the volume resistivity of the conductive films, $L$ is the length of the fiber, $D_F$ and $d_c$ are respectively the diameter of the fiber and thickness of the conductive films. For the longitudinal currents, fiber resistivity will be:

$$R_F^l \propto \rho_v \frac{L}{ND_F d_F} \quad (2),$$

which for longer samples ($L > ND_F$) is much higher than the resistivity for transverse currents. Clearly, when connecting to a fiber using a point probe, fiber resistivity will be dominated by its longitudinal component (2). On the other hand, when covering the fiber outer electrode with a continuous probe (such as metal foil shown in Fig. 4(b)) the effective fiber resistivity will be mostly determined by its transverse component given by equation (1), where $L$ would be the probe length. As predicted by expression (1), in the case of a continuous probe, measured resistivity has to decrease for higher coverage ratios. We would like to note that an alternative and a more practical way of implementing a continuous low resistivity probe for the outside fiber electrode is to spray a conductive paint on the fiber surface.

Fig. 6 demonstrates the effect of the outer electrode coverage ratio by a continuous probe on the fiber capacitance $C_F$ and resistance $R_F$. The measured fiber was 780μm in diameter and 202mm in length containing 20 bilayers of conductive film. We can see that fiber resistance is indeed strongly affected by the electrode coverage ratio while the capacitance has a constant value around 13 nF. As seen from the figure, fiber resistance decreases from 440 kΩ to 120 kΩ as the electrode coverage ratio increases from 5% to 100%. As predicted by expression (1), measured resistivity decreases for higher electrode coverage ratios, however this decrease is not purely inversely proportional to the coverage ratio due to contribution of the longitudinal resistivity. Notably, capacitance of the fibers is not sensitive to the position and size of the electrodes. In these measurements the aluminum foil probe was always placed in the middle of the test fiber.

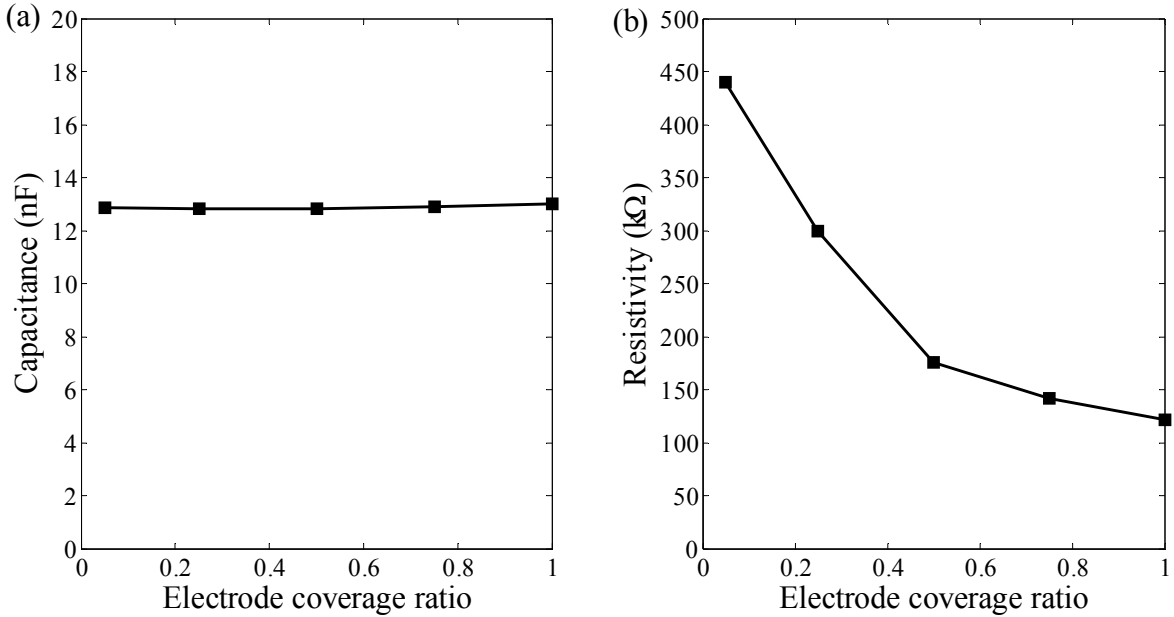

Fig. 6. Effect of the electrode coverage ratio on the capacitance and resistance of a capacitor fiber with diameter of 0.78mm and a length of 202mm.

*B. Effect of the Capacitor Fiber Length*

First we note that capacitance of our fibers can be approximated by that of a parallel plate capacitor with an area equals to the fiber length multiplied by the full length of a spiral electrode, and an inter-plate distance equals to the thickness of an isolating film:

$$C_F \propto \varepsilon_0 \varepsilon \frac{LND_F}{d_i} \quad (3),$$

where $\varepsilon$ is permeability of the isolating films, $\varepsilon_0$ is permeability of the vacuum, $L$ is the fiber length, $N$ is the number of conductive bi-layers in a fiber, $D_F$ and $d_i$ are respectively fiber diameter and thickness of the isolating films. From this it follows that the fiber capacitance is proportional to the fiber length. Moreover, note that fiber capacitance is proportional to the ratio of the fiber diameter to the thickness of isolating layers; if fiber drawing process is homologous, this ratio should stay the same for fiber of any diameter drawn from the same perform. In other words, fiber capacitance per unit length should be the same for the fiber preform as well as for any fiber drawn from such a preform.

In order to study dependence of the capacitor fiber properties as a function of fiber length, we have used two fiber samples that were drawn from the same preform. Sample #1 and sample #2 had outer diameters of 920-980 μm and 720-760 μm respectively. Both samples were drawn form the same preform at speeds in the range 100 mm/min at 180°C. The two samples were then cut into fiber pieces of different lengths ranging between 10 cm and 60 cm, and then wrapped with an aluminum foil with 100% coverage ratio. In Fig. 7(a) we present measured fiber capacitance as a function of fiber length and observe a clear linear dependence. From this data we see that for all the fibers the capacitance per unit length is the around 69 nF/m (inset of Fig. 7(a)), which is very close to the value of 69.5 nF/m measured for the capacitance of the fiber preform. Moreover, as we have demonstrated previously (see equation (1)), for a 100% electrode coverage ratio we expect that capacitor resistivity should decrease inversely proportional with the fiber length. This tendency is clearly observed in Fig. 7(b).

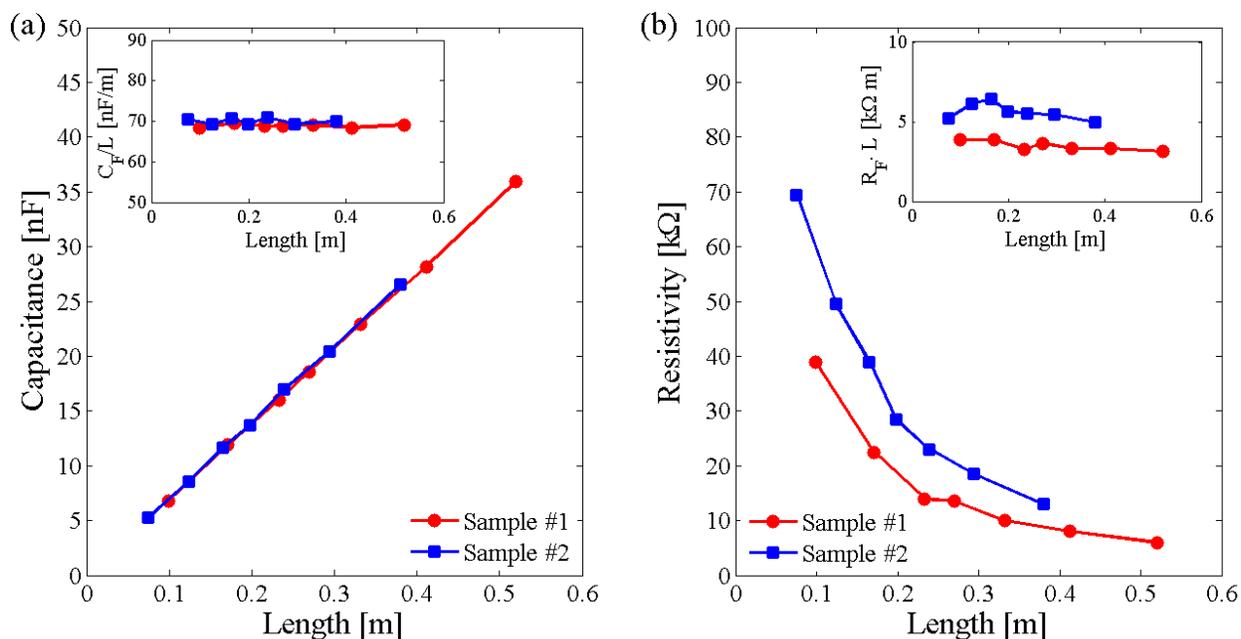

Fig. 7. Dependence of the (a) fiber capacitance and (b) fiber resistivity on fiber length. Red and blue data sets correspond to the two fiber samples of different diameters drawn from the same preform. Insets: dependence of the (a) fiber capacitance per unit length $C_F/L$ and (b) fiber resistivity factor $R_F \cdot L$ on the fiber length.

### C. Effect of the Temperature of Operation on the Fiber Electrical Properties

The effect of the temperature of operation on the electrical properties of a capacitor fiber is presented in Fig. 8 on the example of two particular samples. Sample #1 was 135mm long and had a diameter of 840μm, while sample #2 was 133mm long and had a diameter of 930μm. To control the temperature of two samples they were fixed to a hot plate. Our measurements show that fiber capacitance per unit length remains almost independent of the temperature of operation, while fiber resistivity increases as temperature rises. This result is in good correspondence with the recent reports on positive temperature coefficient [30], [31] for the resistivity of the composites of carbon black and LDPE in the 0°C to 100°C temperature range. The effect of thermal expansion and a consequent increase of the average distance between carbon black particles are thought to be the main reasons for the positive temperature coefficient of such conductive polymer composites. This interesting property promises various applications of capacitor fibers in self-controlled or self-limiting textiles responsive to temperature or heat.

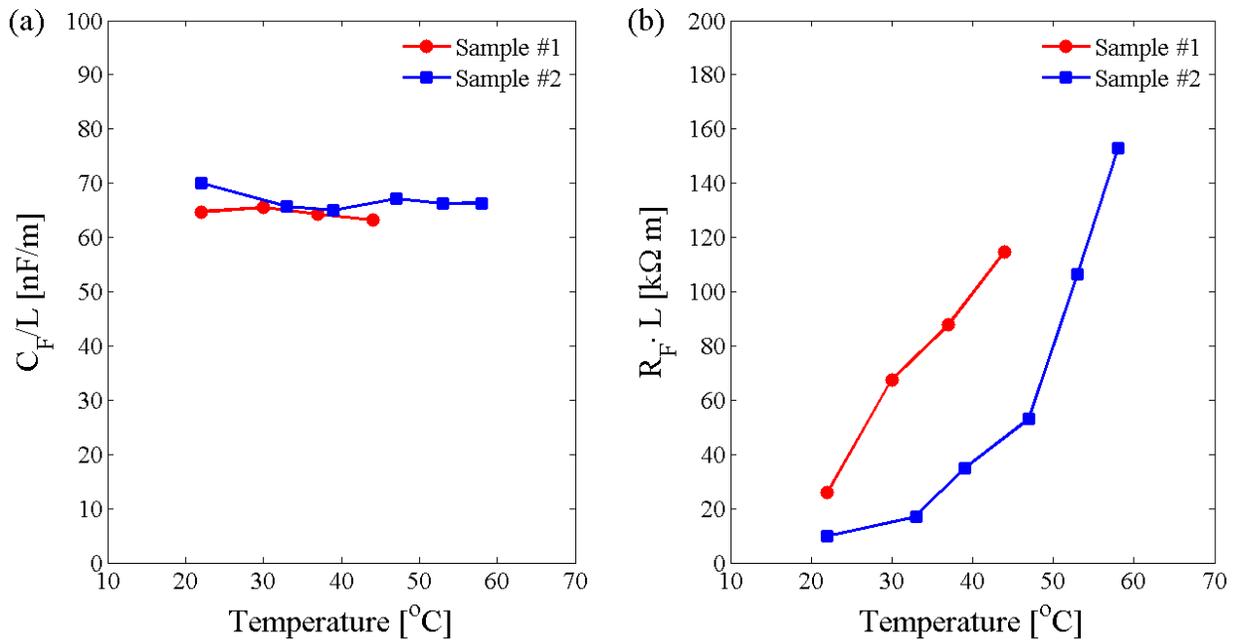

Fig. 8. Effect of the temperature of operation on electrical properties of a capacitor fiber. (a) Capacitance per unit of length $C_F/L$. (b) Resistivity factor $R_F \cdot L$. Sample #1 has a diameter of 840μm and a length of 135mm. Sample #2 has a diameter of 930 m and a length of 136mm.

### D. Effect 0f the Fiber Drawing Parameters

Electrical performance of the capacitor fibers is equally affected by the fiber geometrical parameters and by the fiber material parameters. In this section we show that fiber fabrication parameters such as fiber drawing temperature and fiber drawing speed can have a significant effect on the fiber resistivity, while also somewhat affecting the fiber capacitance. Generally, capacitor fibers presented in this work can be drawn at temperatures in the range from 170˚C to 185˚C with drawing speeds ranging from 100 to 300 mm/min. Fig. 9 presents capacitance per unit length $C_F/L$, and fiber resistivity parameter $R_F \cdot L$ as a functions of the fiber diameter. Several sets of data are presented for the fibers drawn from the same perform at various drawing temperatures and drawing speeds.

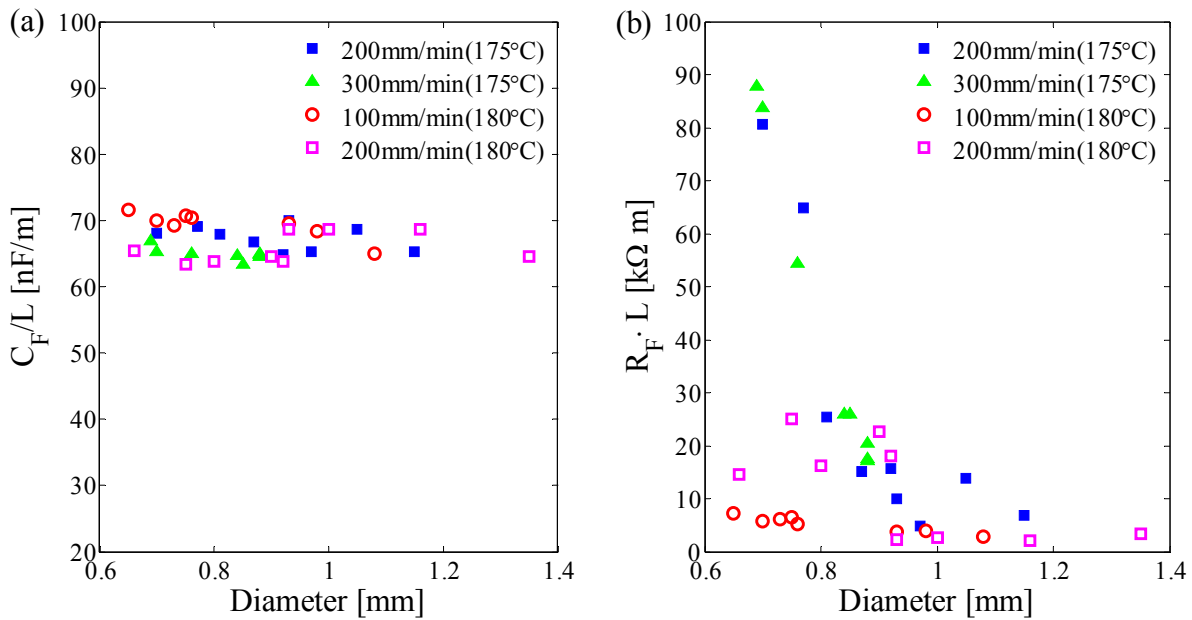

Fig. 9. Electrical properties of the capacitor fibers as a function of fiber drawing parameters. (a) Capacitance per unit length $C_F/L$. (b) Resistivity parameter $R_F \cdot L$. Presented data is for fibers drawn at 175°C and 180°C with three different speeds of 100, 200 and 300 mm/min.

As seen from Fig. 9(a) fiber capacitance $C_F/L$ is largely independent on the fiber diameter and drawing parameters, and equals to that of the fiber perform. In contrast, fiber resistivity parameter $R_F \cdot L$ is significantly affected by the drawing parameters. Although simple analytical model of fiber resistivity (2) predicts that similarly to the fiber capacitance, the resistivity should not vary with the fiber diameter, from Fig. 9(b) we find that this statement is not true. Particularly, from (1) and (3) fiber capacitance per unit length and fiber resistivity parameter are related as:

$$\frac{R_F^t L}{C_F / L} \propto \frac{\rho_v}{\varepsilon_0 \varepsilon} \quad (4)..$$

While fiber capacitance is indeed almost independent of the fiber geometrical and processing parameters, fiber resistivity is strongly influenced by them. This can be rationalised by concluding that bulk resistivity of the carbon black polymer composite can change significantly during drawing procedure from its value in the perform. From Fig. 9 we find that fibers with the lowest resistivity are drawn at higher temperatures and lower drawing speeds. It is in this regime that we could also produce the smallest diameter fibers.

To further validate our observation that drawing at lower temperatures results in higher resistivities we performed a set of stretching experiments at room temperature on the planar conductive films. It was found that the resistivity of the conductive film increases as much as by two orders of magnitude from its original value when the films were stretched unheated to about two times of their length. Similar observation has been reported in literatures for carbon black filled polymers and polymer composites [32], [33]. Conclusions of these experiments can be rationalised by noting that at lower temperatures stretching results in the increase of distance between the individual carbon black particles in the direction of stretching. At higher temperatures, mobility of the carbon black particles in the host polymer matrix increases and they could move easily in the direction of stretching, thus reducing somewhat the inter-particle distance and material resistivity. We, therefore, believe that our experimental observation about lower drawing speeds resulting in lower fiber resisitivities is related to the longer time available at lower drawing speeds for the carbon black particles to move in the direction of stretching during drawing, thus reducing the inter-particle distance and reducing resistiviy of the resultant fibers.

## VII. CONCLUSION

In this work we present novel soft capacitor fibers made of conductive plastics and featuring relatively high capacitance which is 3 to 4 orders magnitude higher than that of a coaxial cable of a comparable diameter. The fibers were fabricated using fiber drawing technique which can be easily scaled up for the industry production. Fibers of various diameters from 650μm to several mm have been demonstrated with a typical capacitance per unit length of 69 nF/m, and a typical resistivity parameter of 5 kΩ·m. It was also demonstrated that during drawing, one or two metallic wire electrodes could be integrated into the fiber structure for the ease of their further connectorization. Developed capacitor fibers are ideally suited for the integration into textile products as they are soft, small diameter, lightweight and do not use liquid electrolytes. Our measurements show that the fiber capacitance is a very stable parameter independent of the fiber diameter, operational temperature and electrical probe structure. In contrast, fiber resistivity has a very strong positive temperature coefficient, it is sensitive to stretching, and strongly dependent on the shape of an electrical probe. We have also demonstrated that while fiber capacitance is proportional to the fiber length, fiber resistivity is inversely proportional to the fiber length (assuming full electrode coverage). We envision the use of thus developed capacitor fibers for various applications in electronic and smart textiles, distributed sensing, and energy storage.

ACKNOWLEDGMENT

This project is supported in part by the NSERC and The Canada Council for the Arts New Media Initiative Karma Chameleon project, as well as Canada Research Chair program. We would also like to thank Prof. J. Berzowska from Concordia University for the fruitful discussions..

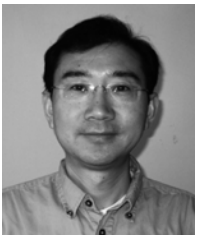
**Dr. Jian Feng Gu** received the B.Eng. degree in dynamic engineering from Huazhong University of Science and Technology, China and M.Eng. degree in petrochemical engineering from Lanzhou University of Technology, China in 1992 and 1995 respectively and Ph.D. degree in chemical engineering from Ecole Polytechnique de Montréal/University de Montréal, Canada in 2008.
From 1995 to 2002, he worked with Shanghai Ship & Shipping Research Institute, China as a process engineer, researcher and a project manager. His Ph.D. thesis was on the modelling and processing of complex fluids including polymers, polymer composites and suspensions. He is currently a postdoctoral fellow in the research group of Professor Maksim Skorobogatiy, a Canada Research Chair in Micro- and Nano-Photonics, at Ecole Polytechnique de Montréal. His current research is concentrated on the development of electronic and piezoelectric fibers and textiles.

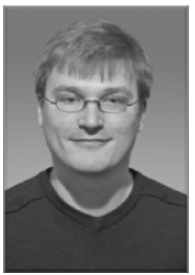
**Dr. Maksim Skorobogatiy** has received his BSc. in physics from Rochester Institute of Technology, USA in 1995, his MSc in Physics from McGill University, Canada in 1997, his MSc in Electrical Engineering and Computer Science from MIT, USA in 2000, and his PhD in Physics from MIT in 2001. During 2001-2003 Dr. Skorobogatiy headed a theory and simulations group at MIT startup OmniGuide Inc., where he was working on commercialization of the hollow photonic bandgap fiber technology for delivery of high power laser beams. Since 2003, Dr. Skorobogatiy is an associate professor and a Canada Research Chair in Micro and Nano-Photonics at Ecole Polytechnique de Montreal / University de Montreal, Canada. His current fields of interest include design and fabrication of photonic crystal fibers for sensing and telecommunications, microstructured fibers for smart textile applkications, as well as the general fields of plasmonics, nano-photonics and nanosciences.